\documentclass[reqno,12pt]{amsart}
\usepackage{amsmath}
\usepackage{amsmath,amssymb,amsthm,graphicx,mathrsfs,url}

\usepackage[top=0.9in,left=0.95in,right=0.95in,bottom=0.9in]{geometry}
\usepackage{subcaption}
\usepackage{amssymb}
\usepackage[dvipsnames]{xcolor}
\newcommand\myshade{85}
\colorlet{mylinkcolor}{violet}
\colorlet{mycitecolor}{YellowOrange}
\colorlet{myurlcolor}{Aquamarine}
\usepackage[unicode=true,pdfusetitle,
bookmarks=true,bookmarksnumbered=false,bookmarksopen=false,
breaklinks=false,pdfborder={0 0 1},backref=false,
linkcolor  = mylinkcolor!\myshade!black,
citecolor  = mycitecolor!\myshade!black,
urlcolor   = myurlcolor!\myshade!black,
colorlinks = true,
]
{hyperref}
\usepackage[nameinlink]{cleveref}

\usepackage{tikz}
\usepackage{autonum} 
\usepackage{dsfont} 
\usepackage{extarrows} 

\usepackage{environ} 
\usepackage{floatrow} 
\usepackage{dsfont}

\newcommand{\QQ}{{\mathcal{Q}}}

\newcommand{\R}{\mathbb{R}}

\newcommand{\NN}{\mathcal{N}}
\newcommand{\p}{{\partial}}

\newcommand{\matrice}[1]{\left[ \begin{matrix}
#1
\end{matrix} \right]}

\newcommand{\Pp}{{\mathbb{P}}}

\newcommand{\epsi}{\varepsilon}

\newcommand{\Span}{{\operatorname{Span}}}
 
\newcommand{\Tr}{{\operatorname{Tr}}}

\newcommand{\lr}[1]{\langle #1 \rangle}

\newcommand{\supp}{\mathrm{supp}}

\newcommand{\Bb}{\mathbb{B}}

\newcommand{\Z}{\mathbb{Z}}

\newcommand{\II}{\mathcal{I}}

\newcommand{\LL}{{\mathcal{L}}}

\newcommand{\tr}[1]{{\textcolor{red}{#1}}}

\newcommand{\C}{\mathbb{C}}

\newcommand{\norm}[1]{\| #1 \|}

\newcommand{\1}{\mathds{1}}
\newcommand{\RR}{\mathcal{R}}
\newcommand{\de}{ \ \mathrel{\stackrel{\makebox[0pt]{\mbox{\normalfont\tiny def}}}{=}} \ }
\newcommand{\Spec}{\operatorname{Spec}}

\numberwithin{equation}{section}

\usepackage[alphabetic,nobysame]{amsrefs}

\newcommand{\alexis}[1]{\textcolor{blue}{#1}}

\newcommand{\jacob}[1]{\textcolor{red}{#1}}

\usepackage{braket}

\newcommand{\eq}[1]{\begin{align}#1\end{align}}
\newcommand{\vf}{{\varphi}}
\newcommand{\br}[1]{\left(#1\right)}
\newcommand{\Id}{\mathds{1}}
\newcommand{\ee}{\operatorname{e}}
\newcommand{\ii}{\operatorname{i}}
\newcommand{\dif}{\operatorname{d}\!}
\newcommand{\ip}[2]{\langle #1, #2 \rangle}
\crefname{section}{\S}{\S}
\crefname{equation}{}{equations}
\Crefname{equation}{Equation}{Equations}
\crefrangelabelformat{equation}{(#3#1#4--#5#2#6)}
\crefmultiformat{equation}{(#2#1#3}{, #2#1#3)}{#2#1#3}{#2#1#3}
\Crefmultiformat{equation}{(#2#1#3}{, #2#1#3)}{#2#1#3}{#2#1#3}

\title{Edge spectrum for truncated $\Z_2$-insulators}
\author{Alexis Drouot}
\address[Alexis Drouot]{University of Washington, Seattle, USA.} 
\email{adrouot@uw.edu}

\author{Jacob Shapiro}

\address[Jacob Shapiro]{Princeton University, Princeton, USA.} 
\email{shapiro@math.princeton.edu}

\author{Xiaowen Zhu}
\address[Xiaowen Zhu]{University of Washington, Seattle, USA.} 
\email{xiaowenz@uw.edu}

\NewEnviron{equations}{%
\eq{\begin{gathered}
  \BODY
\end{gathered}}
}
\theoremstyle{theorem}
\newtheorem{thm}{Theorem}
\newtheorem{lemma}{Lemma}[section]

\newtheorem{proposition}{Proposition}

\theoremstyle{definition}
\newtheorem{definition}{Definition}

\theoremstyle{remark}
\newtheorem{remark}{Remark}[section]

\begin{document}
	\maketitle

   \begin{abstract}
    Fermionic time-reversal-invariant  insulators in two dimension -- class AII in the Kitaev table -- come in two different topological phases. These are characterized by a $\Z_2$-invariant: the Fu--Kane--Mele index. We prove that if two such insulators with different indices occupy regions containing arbitrarily large balls, then the spectrum of the resulting operator fills the bulk spectral gap. Our argument follows a proof by contradiction developed in \cite{DZ23} for quantum Hall systems. It boils down to showing that the $\Z_2$-index can be computed only from bulk information in sufficiently large balls. This is achieved via a result of independent interest: a local trace formula for the $\Z_2$-index.
\end{abstract}

\section{Introduction}

Topological insulators are remarkable phases of matter that are insulating in their bulk but conducting along their boundaries. Mathematically, this is expressed as a spectral-theory result: the spectrum of an interface Hamiltonian between distinct topological phases systematically fills the bulk spectral gap, see e.g. \cite{FGW00,DP02,LT22,DZ23}. A stronger version of this principle is the bulk-edge correspondence, which computes the conductance of the interface in terms of bulk topological invariants. 

The mathematical analysis of these properties has primarily taken place in the context of straight interfaces  \cite{KS02,EG02,EGS05,GP13,BKR,FSSWY20,BC24}. Recently, results regarding the spectrum of curved interface systems have been established for Hall insulators (topological phases characterized by a $\Z$-valued index), see  \cite{LT22,DZ23,DZ24,DSZ24}. These works feature a geometric bulk-edge correspondence formula -- with the emergence of a topological intersection number \cite{DZ24} -- and proofs that the edge spectrum is robust and absolutely continuous \cite{LT22,DZ23,DSZ24}.   

The present work focuses on fermionic time-reversal invariant insulators, models characterized by a $\Z_2$-index \cite{FKM07} that include the Fu--Kane--Mele Hamiltonian and the quantum spin Hall effect \cite{KM05,FKM07,P09}. It extends the gap-filling result of \cite{DZ23} on Hall insulators to $\Z_2$-insulators. The main challenge with $\Z_2$-insulators is the absence of an easily exploitable trace formula for the $\Z_2$-index (for Hall insulators, this is the Kubo formula, see e.g. \cite[equation (1.4)]{EGS05}). 

Our strategy goes around this obstacle by providing a local trace formula for the $\Z_2$-index; specifically, a trace formula valid for Hamiltonians close to a reference system. An important feature of our formula is that closeness is measured in an origin-independent way. We obtain, as applications, that the $\Z_2$-index is locally determined. Because it is also origin-independent, it can be computed only from bulk information in sufficiently large balls. This is an important and seemingly unknown property:  it justifies, for instance, the validity of numerical computations of the $\Z_2$-index in truncated systems. In the present work, it allows us to apply the contradiction strategy of \cite{DZ23} and obtain the filling of the bulk spectral gap when an edge is introduced.

\subsection{Time-reversal-invariant insulators} We review here the mathematical formulation of the Fu--Kane--Mele index \cite{FKM07} due to Schulz--Baldes \cite{SB_2015}; see also \cite{KK18,FSSWY20}.

\begin{definition} A (fermionic) time-reversal map  on $\ell^2(\Z^2,\C^d)$ is an anti-unitary operator\footnote{That is, $\Theta$ is $\C$-antilinear and $\Theta^* \Theta = \Id$, where $\Theta^*$ is the $\C$-antilinear operator implicitly defined by $\langle \Theta u, v\rangle = \overline{\langle u, \Theta^* v\rangle}$.} $\Theta$ such that $\Theta^2 = - \Id$ and $[\Theta,x_j] = 0$ for $j=1,2$ (where $x_j$ denotes the position operator in the $j$-th direction). 
\end{definition}

\begin{definition} Let $H$ on $\ell^2(\Z^2,\C^d)$ be a bounded, selfadjoint Hamiltonian. We say that $H$~is:
\begin{itemize}
    \item \textit{local} if its (integral) kernel $\Set{H(x,y)}_{x,y\in\Z^2}$ satisfies, for some $\nu > 0$,
\begin{align}
    | H(x,y)| \leq \nu^{-1} e^{-2\nu|x-y|}, \qquad x,y \in \Z^2. 
\end{align}
    \item \textit{insulating at energy $E_F$} if $E_F$ is not in the spectrum $\Spec(H)$ of $H$.
    \item \textit{$\Theta$-invariant} if $[H,\Theta] = 0$.
\end{itemize}
\end{definition}

\begin{definition}  The $\Z_2$-index of a local, $\Theta$-invariant Hamiltonian $H$ insulating at energy $E_F$ is
\eq{\label{eq:FKM index}
    \II(H) \de \dim \ker(PLP+P^\perp) \mod 2 \qquad\in \Z_2\,, \qquad \text{where}
} 
\begin{itemize}
    \item $P = \1_{(-\infty,E_F)}(H)$ is the spectral projection below energy $E_F$ and $P^\perp = \Id-P$;
    \item $L$ is the Laughlin flux insertion operator, i.e. the multiplication operator by \eq{x\mapsto \ee^{\ii\arg(x_1+ix_2)}\,.} 
\end{itemize}
\end{definition}

The quantity $\II(H)$ is well-defined because the kernel in \eqref{eq:FKM index} has finite dimension. Indeed, because $[P, L]$ is compact (see \Cref{lem:2d} below), $PLP+P^\perp$ is Fredholm as an invertible modulo compact (Atkinson's theorem):
\begin{equations}
    (PL^*P+P^\perp)(PLP+P^\perp) = PL^*PLP + P^\perp = P + P^\perp + PL^*[P,L] P = \Id + PL^*[P,L] P, \\
    (PLP+P^\perp)(PL^*P+P^\perp) = PLPL^*P + P^\perp = P + P^\perp + PL[P,L^*] P = \Id - P[L,P]L^* P.
\end{equations}

The integer $\dim\ker(F) \mod 2$ in \cref{eq:FKM index} was first proposed as a topological index for skew-adjoint Fredholm operators $F$ by Atiyah and Singer in \cite{Atiyah1969}. They established that (in the present notation), for Fredholm $F$ satisfying \eq{\label{eq:Theta-odd condition}
    \Theta F \Theta = - F^\ast, 
}
$\dim \ker(F) \mod 2$ is invariant w.r.t norm continuous and compact perturbations which also obey \cref{eq:Theta-odd condition}; in our notation, $F = PLP+P^\perp$ and $\II(H) = \dim \ker(F) \mod 2$. We review the properties of $\II(H)$ in \S\ref{sec:3}. Note that despite the notation, $\II(H)$ is a function of $E_F$; but because it specifically depends on $P$ only, it is constant for energies in the spectral gap.



\subsection{Interface systems} We now introduce \textit{interface Hamiltonians.} 

\begin{definition}\label{def:edge} Let $H$ be a local, $\Theta$-invariant Hamiltonian $H$ on $\ell^2(\Z^2,\C^d)$. We say that $H$ \textit{models an interface system at energy} $E_F$ if there exists $\Omega \subset \R^2$ and two local, $\Theta$-invariant Hamiltonians $H_+, H_-$ such that:
\begin{enumerate}
    \item[(i)] $\Omega$ and $\Omega^c$ contain arbitrarily large balls: 
    \eq{
        \forall R > 0, \ \exists a_+, a_- \in \R^2, \quad \Bb_R(a_+) \subset \Omega, \ \Bb_R(a_-) \subset \Omega^c.
    }
    \item[(ii)] The Hamiltonians $H_+, H_-$ are insulating at energy $E_F$ and $\II(H_+) \neq \II(H_-)$.
    \item[(iii)] The kernel of the edge operator \eq{\label{eq:edge operator}E \de H - \1_\Omega H_+ \1_\Omega - \1_{\Omega^c} H_- \1_{\Omega^c}} decays away from the interface, where $\Id_S$ denotes the multiplication operator by the indicator function of a set $S \subset \Z^2$. That is, the kernel of $E$ satisfies, for some $\nu > 0$:
    \eq{\label{eq:decay of edge operator away from interface}
        |E(x,y)|\leq \nu^{-1} e^{-\nu d(x,\p \Omega)}, \qquad x \in \Z^2.
    }
\end{enumerate}
\end{definition}

One way to realize such an interface system is to simply fix $\Omega$ such that (i) holds, $H_+$ and $H_-$ such that (ii) holds, and to simply set $H = \1_\Omega H_+ \1_\Omega + \1_{\Omega^c} H_- \1_{\Omega^c}$. The resulting operator models the junction of two distinct topological phases along the interface $\p \Omega$. We refer to $H_+$ and $H_-$ as the bulk systems and to $H$ as the (interface) edge system. While $E_F$ is an insulating energy for the bulk systems, an interface system is always conducting at energy $E_F$:

\begin{thm}\label{thm:4} Assume that $H$ models an interface system at energy $E_F$ in the sense of \Cref{def:edge} (in particular, $\Omega$, $\Omega^c$ contain arbitrarily large balls and $\II(H_+) \neq \II(H_-)$). Then $E_F \in \Spec(H)$. 
\end{thm}

When $\Omega$ is a half-plane, \Cref{thm:4} is a weak version of the bulk-edge correspondence \cite[Theorem 2.11]{FSSWY20}: it predicts that an interface between two topologically distinct phases cannot itself be insulating. When $\Omega$ has a more complex geometry, \Cref{thm:4} is new and analogous to \cite[Theorem 1]{DZ23} for quantum Hall systems. In a forthcoming paper, we will show that $H$ has absolutely continuous spectrum near $E_F$ and that the $\Z_2$-edge index is given by the difference of the two $\Z_2$-bulk indices times an intersection number. This will extend the works \cite{DZ23, DSZ24} to $\Z_2$-insulators.

For \Cref{thm:4} to hold, a condition on $\Omega$ is necessary. In \S\ref{sec:5}, we show that if $\Omega$ lies in a strip of arbitrary but finite width, then $E_F$ may lie in a spectral gap of $H$.

\subsection{Strategy} 
The main tool in the proof of \Cref{thm:4} essentially states that $\Z_2$-index of an insulating Hamiltonian $H$ can be computed from only approximate knowledge of $H$ in a sufficiently large ball centered anywhere:

\begin{proposition}\label{prop:5}
        Let $H_1,H_2$ be local, $\Theta$-invariant Hamiltonians, insulating at energy $E_F$. There exist $\epsi, R>0$ such that if for some $a \in \Z^2$, 
\begin{equations}
    \big\| \1_{\Bb_R(a)} (H_1-H_2) \1_{\Bb_R(a)} \big\| \leq \epsi,
\end{equations}
then $\II(H_1) = \II(H_2)$.
\end{proposition}

To the best of our knowledge,  \Cref{prop:5} is a new result for $\Z_2$-insulators. We then prove \Cref{thm:4} following the contradiction argument of \cite{DZ23}: if $H$ were insulating at energy $E_F$, then its $\Z_2$-index $\II(H)$ would be well-defined. Because $H$ and $H_+$ essentially coincide in suitably chosen large balls,  we would have moreover $\II(H) = \II(H_+)$. But likewise, $\II(H) = \II(H_-)$; this contradicts $\II(H_-) \neq \II(H_+)$. Therefore, $H$ is insulating at energy $E_F$. See \cref{sec:4} for the details of the proof.

\subsection{Trace formulas for the $\Z_2$-index and the proof of \Cref{prop:5}.} As in the case of quantum Hall systems \cite{DZ23}, the core of the paper consists of proving \Cref{prop:5}. In \cite{DZ23}, the analogous result is based on an analysis of the Kubo double-commutator trace formula for the bulk index:
\eq{
    \sigma(H) = -2\pi \ii \cdot \Tr \left( P \big[ [\1_{x_1 > 0},P], [\1_{x_2 > 0},P]\big]\right).
}
But for time-reversal-invariant insulators, there is no known convenient universal trace formula for the $\Z_2$-index. 

Our inspiration comes from a paper due to Katsura--Koma \cite{KK18}. Given a $\Theta$-invariant Hamiltonians $H$, insulating at energy $E_F$, define
\eq{
    B := |P^\perp LP|^6.
}
We will see in the proof of \Cref{prop:5} that, due to the power $6$, $B$ is a trace-class operator. If $\gamma_B$ is a ($B$-dependent) complex contour enclosing $1$ but not $0$, such that $B$ has no eigenvalues on $\gamma_B$, we will moreover justify in \eqref{eq:5a} that 
\eq{\label{eq:4j}
    \II(H) = \Tr \br{\oint_{\gamma_B} \br{B-z\Id}^{-1} \dfrac{\dif{z}}{2\pi \ii}} \mod 2.
}
There are two main ideas behind \eqref{eq:4j}. The first one is that due to some standard algebraic manipulations, $\II(H)$ is equal to the parity of the geometric multiplicity of $1$ as an eigenvalue of $B$. The second one is that while the right hand side of \eqref{eq:4j} counts the sum of all multiplicities of eigenvalues of $B$ enclosed by $\gamma_B$, these multiplicities are all even except for that of $1$. This is due to time-reversal invariance of $H$. In particular, taking modulo $2$ produces $\II(H)$.

There is no contour $\gamma$ that works for all insulators in \cref{eq:4j}, since it must be chosen in a way that avoids the eigenvalues of $B$. Instead, given two insulators $H_1, H_2$, we can construct a contour $\gamma$ that works for both assuming they are sufficiently "close" in a suitable sense. This allows us to express indices $\II(H_1)$, $\II(H_2)$ using \cref{eq:4j} over the same contour $\gamma$. Hence we get
\eq{\label{eq:4k}
    \II(H_2)-\II(H_1) = \Tr \br{\oint_\gamma \br{B_2-z\Id}^{-1} (H_1-H_2) \br{B_1-z\Id}^{-1} \dfrac{\dif{z}}{2\pi \ii}} \mod 2,
}
where $B_j = |P_j^\perp L P_j|^6$, $P_j = \Id_{(-\infty, E_F)}(H_j)$, $j = 1,2$. The proof of \Cref{prop:5} follows then from a technical analysis of \cref{eq:4k}.

\begin{remark}
    The definition \cref{eq:FKM index} of $\Z_2$-indices remains valid in the presence of a mobility gap of $H$ (i.e. the gap fills with Anderson localized states; see \cite{Bols2023} for context and proper definition). However, the scheme presented here to develop a local perturbation theory of the index uses crucially the presence of an actual spectral gap. We postpone an extension of \Cref{prop:5} in the mobility gap regime to a future paper.
\end{remark}

\subsection{Notation}
\begin{itemize}
\item Given $x\in \R^2$ or $\Z^2$, we denote $\langle x\rangle 
 = (1 + |x|^2)^{1/2}$ and we sometimes treat $x$ as a complex number $x = x_1 + ix_2$ without ambiguity. 
\item For a bounded operator $A$, $\| A \|$ denotes the operator norm.
    \item For a projection $P$, we use the shorthand $\Pp L = PLP+P^\perp$.
    \item $\LL^p$ is the Schatten-p class, i.e. the set of bounded operators $A$ such that $|A|^p$ is trace-class. We denote by $\| \cdot \|_1$ the trace-class norm.
    \item For a bounded operator $A$, $|A|^2 = A^*A$.
    \item For a bounded operator $A$ and $\lambda \in \C$, $m_\lambda(A)$ is the geometric multiplicity of $\lambda$ as an eigenvalue of $A$, i.e., $\dim\ker (A-\lambda\Id)$.
    \item $\Bb_R(a)$ denotes the Euclidean ball of radius $R$ centered at $a$, a subset of $\Z^2$.
    \item $\1_\Omega$ is the indicator function of a set $\Omega$.
\end{itemize}

\subsection*{Acknowledgement} We gratefully acknowledge support from the National Science Foundation DMS 2054589 (AD) and the Pacific Institute for the Mathematical Sciences (XZ). The contents of this work are solely the responsibility of the authors and do not necessarily represent the official views of PIMS.

\section{Center-independence}\label{sec:2}

We start with a lemma which implies $\II(H)$ is well-defined:

\begin{lemma}\label{lem:2d} Let $H$ be a local, $\Theta$-invariant Hamiltonian, insulating at energy $E_F$. Then $[P,L] \in \LL^3$. \end{lemma}

\begin{proof} We first note that $L$ is the multiplication operator on $\ell^2(\Z^2,\C^d)$:
\eq{
    (L\psi)(x) = f(x)\psi(x), \qquad f(x) := \dfrac{x}{|x|}, \qquad x\in \Z^2. 
}
Note that there exists a constant $D>0$ such that 
\eq{\label{eq:4l}
    \left| f(x) - f(y) \right| \leq D|x-y| \lr{y}^{-1}, \qquad x, y\in\R^2. 
} 
Meanwhile, $P = \1_{(-\infty, E_F)}(H)$ is the spectral projection of a local insulator $H$. By writing the contour integral representation for the projections, and estimating the resolvents using the Combes-Thomas estimate \cite[\S10.3]{AizenmanWarzel2016}, we have that $P$ is also local and satisfies
\eq{
|P(x, y)|\leq \nu^{-1} e^{-\nu|x - y|};
}
for more details see
e.g. \cite[Lemma 3.2]{DZ23}.
As a result, the kernel of $T = [P,L]$ satisfies 
\eq{
 \big| T(x,y) \big| =
    \big| P(x,y) \big(f(x) - f(y)\big) \big| \leq \nu^{-1} e^{-\nu |x-y|} \big| f(x)-f(y) \big| \leq D\nu^{-1} |x-y| e^{-\nu |x-y|} \lr{y}^{-1}. 
}
In particular, 
\eq{
    \sum_{b \in \Z^2} \left( \sum_{x \in \Z^2} \big| T(x+b,x) \big|^3 \right)^{1/3} \leq D\nu^{-1}  \left(\sum_{b \in \Z^2} |b| e^{-\nu |b|}\right) \cdot  \left(\sum_{x\in \Z^2} \lr{x}^{-3} \right)^{1/3} < \infty.
}
By \cite[Lemma 1]{Aizenman_Graf_1998}, $T \in \LL^3$ as claimed. 
\end{proof}

Note that by direct computation (see also \cite{ASS94b}), we have
\eq{\label{eq:4n}
    |\Pp L|^2 = \Id - |A|^2, \qquad \text{where~} A := P^\perp L P = P^\perp [L,P].
}
We see that $\Pp L$ has a finite-dimensional kernel if and only if $1$ is an eigenvalue with a finite multiplicity of $|A|^2$. The latter holds because by \Cref{lem:2d}, $[L, P]$ is compact; hence $A$ is compact. In particular, this implies that $\II(H)$ is well-defined with $\II(H) = m_1(|A|^2) \mod 2$.

For a local, time-reversal invariant Hamiltonian $H$, insulating at energy $E_F$, we introduce a shifted version of the $\Z_2$-index:
\eq{\label{eq:4p}
    \II^{(a)}(H) \de \dim \ker \br{ \Pp L^{(a)} }, \qquad a \in \R^2.
}
In \cref{eq:4p}, $L^{(a)}$ is the Laughlin flux centered at $a$:
\eq{
    L^{(a)} \de \ee^{\ii \arg(x_1+\ii x_2-a_1-\ii a_2)}.
}
One key property of $\Z_2$-index (and similarly, Hall conductance in \cite{DZ23}) is center independence:  $\II^{(a)}(H)$ and $\II(H)$ are equal.

\begin{lemma}\label{lem:6} If $H$ is a local, time-reversal invariant Hamiltonian $H$, insulating at energy $E_F$, then for all $a \in \R^2$, $\II(H) = \II^{(a)}(H)$.     
\end{lemma}

\begin{proof} If one can show that $\Pp L - \Pp L^{(a)}$ is compact, then using invariance of indices of $\Theta$-invariant operators under compact perturbations \cite[Theorem A.14]{FSSWY20}, we will have -- with equalities below valid modulo $2$:
\eq{
    \II^{(a)}(H) = \dim \ker \Pp L^{(a)} =  \dim \ker \big( \Pp L + \Pp (L-L^{(a)}) \big) = \dim \ker \Pp L = \II(H).
    }

Therefore, it suffices to prove that
\eq{
    \Pp L - \Pp L_a = PLP - PL^{(a)}P = P(L-L^{(a)})P 
}
is compact. As $P$ is a bounded operator, it suffices to show that $L-L^{(a)}$ is compact. This is the multiplication operator by
\eq{
    x\mapsto f(x) - f(x-a), \qquad f(x) := \dfrac{x_1+ix_2}{|x|}.
}
By \eqref{eq:4l}, for some $D>0$, we have 
\eq{
\Tr(|L - L^{(a)}|) \leq \sum\limits_{x\in \Z^2} |f(x) - f(x - a)| \leq D\sum_{x\in \Z^2} \frac{|a|}{\langle x\rangle}<+\infty.
} 
Hence $L-L^{(a)}$ is compact. This completes the proof. 
\end{proof}

    \section{Local-determinancy}\label{sec:3}

In \S\ref{sec:2}, we showed that $\II(H) = m_1(|A|^2) \mod 2$. We push the argument further by showing that the parity of the multiplicity $1$ eigenvalue of $|A|^2$ does not change under suitable perturbations:

    \begin{lemma}\label{lem:local formula for Z_2}
        Let $H_1, H_2$ be two local, $\Theta$-invariant Hamiltonians, insulating at energy $E_F$. Let $P_1, P_2$ be the Fermi projections at $E_F$ and $A_j=P_j^\perp L P_j, \  B_j := |A_j|^6, \ j=1,2$. If
        \eq{\label{eq:4e}
            \| B_1-B_2 \|_1 \leq 2^{-8} (\| B_1\|_1 +1)^{-2},
        }
        then $\II(H_1) = \II(H_2)$. 
    \end{lemma}
    
\begin{figure}
    \centering
    \includegraphics[]{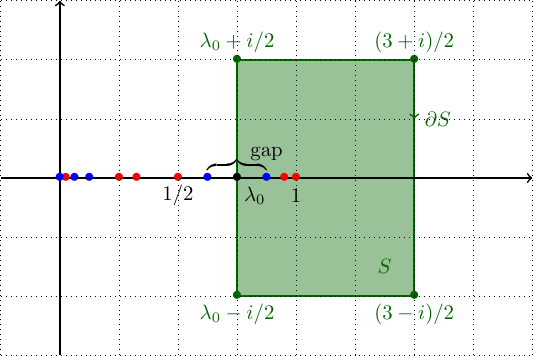}
    \caption{The eigenvalues of $B_1,B_2$ (blue, red) and the contour $\p S$ (green). It is a rectangle passing through the point $\lambda_0$. Its distance to the spectra of $B_1, B_2$ is half the length of the gap, hence at least $2^{-4} (\|B_1\|+1)^{-1}$. It length is at most $4$. }
    \label{fig:1}
\end{figure}

    \begin{proof}
        \textit{Step 1.} First we claim the non-zero eigenspaces of $|\Pp 
        L|^2$ are even dimensional (see e.g. \cite[Theorem A.2]{CS23}). Indeed, $\Pp L$ is $\Theta$-odd, i.e. $\Theta \br{\Pp L} \Theta = - \br{\Pp L}^\ast$. This uses that because $\Theta$ is anti-unitary and commutes with $x$, $U\Theta = \Theta U^*$; and that because $H$ is $\Theta$-invariant, $[P, \Theta] = 0$. Now if $|\Pp L|^2 \psi = \lambda \psi$ for some $\psi\neq0$ and $\lambda > 0$, then $\vf = \Theta \Pp L \psi$ is a linearly-independent eigenvector of $|\Pp L|^2$ with eigenvalue $\lambda$. Indeed,
        \eq{
            |\Pp L|^2 \vf = \Theta \Pp L \Pp L^\ast \Pp L \psi = \Theta \Pp L \lambda \psi = \lambda \vf.
        } Moreover, it is perpendicular to $\psi$:
        \eq{
            \ip{\psi}{\vf} = \ip{\psi}{\Theta \Pp L \psi} = \ip{\Theta \Theta \Pp L \psi}{\Theta \psi} = -\ip{ \Pp L \psi}{\Theta \psi} = -\ip{  \psi}{\Pp L^\ast \Theta \psi} = -\ip{  \psi}{\Theta\Pp L  \psi} = - \ip{\vf}{\psi}.
        }
        Because  $\Theta \Pp L \vf = -\lambda \psi$, the span of $\psi,\vf$ is invariant under the action of $\Theta \Pp L$. Let now $\eta$ be orthogonal to  $\Span(\psi,\vf)$. Then $\Theta \Pp L \eta$ is also orthogonal to that span. Indeed, if $\chi \in \Span(\psi,\vf)$, then $\eta$ and $\chi$ are orthogonal and  \eq{
            \ip{\chi}{\Theta \Pp L \eta} =  \ip{\Theta \Theta \Pp L \eta}{\Theta \chi} = -\ip{ \Pp L \eta}{\Theta \chi} = -\ip{  \eta}{\Pp L^\ast\Theta \chi} = -\ip{  \eta}{\Theta\Pp L \chi} = 0
        } because $\Theta \Pp L$ leaves the span invariant. Hence the non-zero eigenspaces of $|\Pp L|^2$ are indeed even dimensional. 

        
        \textit{Step 2.} From \cref{eq:4n} and the fact that $[L,P_j]$ is compact (see \Cref{lem:2d}), $\Spec(|A_j|^2)\setminus \{0\}$ consists of discrete eigenvalues with finite multiplicity. By \textit{Step 1}, all these eigenvalues except $1$ have even multiplicities. 
        
        In particular, $\Spec(B_j) \setminus \{0\} = \Spec(|A_j|^{6})\setminus \{0\}$ also consists of discrete eigenvalues with even multiplicity, except possibly $1$. Because $x\mapsto x^3$ is bijective,
        \eq{
            m_1(B_j) = m_1(|A_j|^2) = m_0(|\Pp_j L|^2) = \II(H_j) \mod 2.
        }

        \textit{Step 3.} Let $S$ be a closed rectangle in the complex plane such that $B_j$ has no eigenvalues on $\p S$, with $0 \notin S$, $1 \in S$. Then $B_j$ has finitely many eigenvalues in $S$ and we have:
        \eq{
            \Tr \left(\dfrac{1}{2\pi} \oint_{\p S} \br{B_j-z\Id}^{-1} \dif{z} \right) = \sum_{\lambda \in S} m_\lambda(B_j). 
        }
        By \textit{Step 2}, all terms $m_\lambda(B_j)$ except $m_1(B_j)$ are even. We obtain:
        \eq{\label{eq:5a}
            \II(H_j) = m_1(B_j) = \sum_{\lambda \in S} m_\lambda(B_j) \mod 2= \Tr\br{ \dfrac{1}{2\pi \ii} \oint_{\p S} \br{B_j-z\Id}^{-1} \dif{z} } \mod 2. 
        }
        In particular, by identifying $\II(H_j)$ with the element of $\{0,1\}$ with same parity: 
        \begin{equations}\label{eq:4c}
            \big|\II(H_1)-\II(H_2) \big| \leq \Delta, \qquad \text{with} 
            \\
        \Delta \de \left| \Tr \left(\dfrac{1}{2\pi \ii} \oint_{\p S} (B_1-z\Id)^{-1} \dif{z} \right) - \Tr \left(\dfrac{1}{2\pi \ii} \oint_{\p S} (B_2-z\Id)^{-1} \dif{z} \right) \right|\,.
        \end{equations}

        \textit{Step 4.} We now estimate $\Delta$:
        \begin{equations}
            \Delta \leq 
            \left\| \dfrac{1}{2\pi \ii} \oint_{\p S} (B_1-z\Id)^{-1}  (B_1 - B_2) (B_2-z\Id)^{-1} \dif{z} \right\|_1
            \\ 
            \leq \dfrac{|\p S| \cdot \| B_1 - B_2 \|_1}{2\pi \cdot d(\Spec(B_1),\p S) \cdot d(\Spec(B_2),\p S)}.    \label{eq:4d}         
        \end{equations}

        Consider the set $\Sigma_j = \Spec(B_j) \cap \left[\frac12,\infty\right)$. For $\lambda \in \Sigma_j$, $2\lambda \geq 1$ and therefore:
        \eq{
            \# \Sigma_j = \sum_{\lambda \in \Sigma_j} 1 \leq \sum_{\lambda \in \Sigma_j} 2 \lambda \leq 2 \sum_{\lambda \in \Spec(|A_j|^{6})} = 2\| B_j \|_{\LL^1}.
        }
       Note by \cref{eq:4e}, we have $\| B_2\|_1 \leq \|B_1\|_1 + 1/2$. Hence we obtain
        \eq{
            \#(\Sigma_1 \cup \Sigma_2) \leq 2 (\| B_1 \|_{\LL^1} + \| B_2 \|_{\LL^1}) \leq 4(\|B_1\|+1).  
        }
        It follows that there exists an interval contained in $[1/2,1]$ of length $2^{-3} (\|B_1\|+1)^{-1}$ that contains no eigenvalues of $A_1$ nor $A_2$. Let $\lambda_0$ be the midpoint of this interval and $\p S$ be the rectangle such that $\lambda_0 \in \p S$, see \Cref{fig:1}. We have $|\p S| \leq 4$ and by \cref{eq:4d},
        \eq{
            \Delta \leq 
            \dfrac{4\| B_1 - B_2 \|_1}{2\pi \cdot 2^{-8} (\|B_1\|+1)^{-2}} 
            \leq \dfrac{2^2 \cdot 2^{-8}   \br{\|B_1\|_1+1}^{-2}}{2\pi \cdot 2^{-8}   \br{\|B_1\|_1+1}^{-2} } < 1.
        }
        This implies that $\II(H_1)=\II(H_2)$. 
    \end{proof}

\begin{remark} The same logic could have been applied to the index of the integer quantum Hall effect (IQHE) to re-derive \cite[Proposition 1]{DZ23} here. Specifically, if two IQHE Hamiltonians $H_1,H_2$ obey \cref{eq:4e} (with the same notation as above) then they have the same Hall conductance. Indeed, the Hall conductance associated to $P$ equals
    \eq{\label{eq:5b}
        \operatorname{index}(\Pp L) = m_0(|\Pp L|^2) - m_0(|\Pp L^\ast|^2) = m_1(|A|^2) - m_1(|A^\ast|^2),
    }  see for instance \cite{ASS94}.  We may extend \eqref{eq:5b} to 
    \begin{equation}
        \operatorname{index}(\Pp L) = \sum_{\lambda \in S} m_\lambda(|A|^2) - m_\lambda(|A^\ast|^2),
    \end{equation}
    where $S$ is e.g. a square enclosing $1$ but not $0$. This relies on a one-to-one correspondence between the eigenspaces of $|A|^2$ and of $|A^\ast|^2$ with eigenvalues different from $1$. 
    
    Taking powers of these operators preserves the multiplicities (as the operators are positive so powers are bijective). By picking a square $S$ whose boundary contain no eigenvalues of $H_1$ nor $H_2$, we obtain \eq{
    \operatorname{index}(\Pp_1 L) - \operatorname{index}(\Pp_2 L) &= \dfrac{1}{2\pi \ii} \oint_{\p S} (B_1-z\Id)^{-1}  (B_1 - B_2) (B_2-z\Id)^{-1} \dif{z} -\\
    &-\dfrac{1}{2\pi \ii} \oint_{\p S} (Y_1-z\Id)^{-1}  (Y_1 - Y_2) (Y_2-z\Id)^{-1} \dif{z}
    } as long as $p$ is such that $B_j = |A_j|^{2p}$ and $Y_j = |A_j^\ast|^{2p}$ are trace-class and $\partial S$ encircles $1$ but not zero, and passes through no eigenvalues of $B_j,Y_j$ for $j=1,2$.
\end{remark}

We are now ready for the proof of
\begin{proposition}[Restatement of \Cref{prop:5}]
        Let $H_1,H_2$ be local, $\Theta$-invariant Hamiltonians, insulating at energy $E_F$. There exist $\epsi, R>0$ such that if for some $a \in \Z^2$, 
\begin{equations}\label{eq:4f}
    \big\| \1_{\Bb_R(a)} (H_1-H_2) \1_{\Bb_R(a)} \big\| \leq \epsi
\end{equations}
then $\II(H_1) = \II(H_2)$.
\end{proposition}

\begin{proof}[Proof of \Cref{prop:5}] \textit{Step 1.} We note that 
\begin{align}
    B_j & = P_j L^* P_j^\perp L P_j L^* P_j^\perp L P_j L^* P_j^\perp L P_j 
    \\
    & = P_j [L^*, P_j^\perp] [L, P_j] [L^*, P_j^\perp] [L, P_j] [L^*, P_j^\perp] [L,P_j]
    \\
    \label{eq:4m}
    & = -P_j [L^*, P_j] [L, P_j] [L^*, P_j] [L, P_j] [L^*, P_j] [L,P_j].
\end{align}

\textit{Step 2.} We recall that $T_j = [L, P_j] \in \LL^3$, see \Cref{lem:2d}. In particular, triple products of operators $T_j$ or $T_\ell^*$ are trace-class. Therefore, using the multilinear expression \cref{eq:4m} of $B_j$, we deduce that we can write 
\eq{\label{eq:4i}
B_1-B_2 = 
    K_1 (P_1-P_2) L_1 + L_2 (P_1-P_2) K_2,
}
where $K_1$ and $K_2$ are trace-class, and $L_1, L_2$ are bounded. 

\textit{Step 3.} For $r > 0$, we have:
\begin{align}
    K_1 (P_1-P_2) L_1 & = K_1 \1_{\Bb_r(0)} (P_1-P_2) \1_{\Bb_r(0)} L_1 + K_1 \1_{\Bb_r(0)^c} (P_1-P_2) L_1  
    \\ 
    & + K_1 \1_{\Bb_r(0)} (P_1-P_2) \1_{\Bb_r(0)^c} L_1
    \\
    & = K_1 \1_{\Bb_r(0)} (P_1-P_2) \1_{\Bb_r(0)} L_1 + K_1 \1_{\Bb_r(0)^c} (P_1-P_2) L_1 
    \\
    & + K_1 (P_1-P_2) \1_{\Bb_r(0)^c} L_1 - K_1 \1_{\Bb_r(0)^c} (P_1-P_2) \1_{\Bb_r(0)^c} L_1.
\end{align}
So, for a constant $C$ independent of $r$,
\begin{equations}\label{eq:4h}
    \big\| K_1 (P_1-P_2) L_1 \big\|_1  
    \\
    \leq C \left( \big\| \1_{\Bb_r(0)} (P_1-P_2) \1_{\Bb_r(0)} \big\| + \big\| K_1 \1_{\Bb_r(0)^c} \big\|_1 + \big\| K_1 (P_1-P_2) \1_{\Bb_r(0)^c} \big\|_1\right).
\end{equations}
Set $\epsilon = 2^{-10} (1+\| B_1 \|_1)^{-2}$. As $r \rightarrow \infty$, $\1_{\Bb_r(0)^c} \rightarrow 0$ strongly so, because $K_1$ is trace-class, $K_1 \1_{\Bb_r(0)^c} \rightarrow 0$ in $\LL^1$. Likewise, $K_1 (P_1-P_2) \1_{\Bb_r(0)^c} \rightarrow 0$ in $\LL^1$. In particular, coming back to \cref{eq:4h}, there exists $r > 0$ such that
\begin{equations} \label{eq:4g}
    \big\| K_1 (P_1-P_2) L_1 \big\|_1 \leq C \big\| \1_{\Bb_r(0)} (P_1-P_2) \1_{\Bb_r(0)} \big\| + \epsilon. 
\end{equations}

\textit{Step 4.} Set now $R = 2r$, $\epsi = \epsilon/(2C)$, and assume that $H_1, H_2$ satisfy 
\eq{
    \big\| \1_{\Bb_R(0)} (H_1-H_2) \1_{\Bb_R(0)} \big\| \leq \epsi.
}
After possibly increasing the value of $r$ (note that this does not affect the validity of \cref{eq:4g}), we have 
\eq{
    \big\| \1_{\Bb_r(0)} (P_1-P_2) \1_{\Bb_r(0)} \big\| \leq 2\epsi = \epsilon/C.
} This follows by writing the contour integral representation for the projections, and estimating the resolvents using the Combes-Thomas estimate \cite[Section 10.3]{AizenmanWarzel2016}; for more details see e.g. \cite[Lemma 3.2]{DZ23}. We thus deduce that
\eq{
    \big\| K_1 (P_1-P_2) L_1 \big\|_1 \leq 2 \epsilon.
}
Applying the same argument to $L_2 (P_1-P_2) K_2$ and coming back to \cref{eq:4i}, we deduce that
\eq{
    \| B_1 - B_2 \| \leq 4 \epsilon = 2^{-8} (1+\| B_1 \|_1)^{-2}.
}
In particular, \Cref{lem:local formula for Z_2} implies that $\II(H_1) = \II(H_2)$. This completes the proof when $a = 0$.

\textit{Step 5.} If now $H_1, H_2$ satisfy \cref{eq:4f} for some $a \neq 0$ and the values of $\epsi, R$ given in Step 4, then the shifts $H_1^{(a)}, H_2^{(a)}$ of $H_1, H_2$ by $a$ (i.e. the operators $H_j^{(a)} = T_a H_j T_a^*$, where $T_a\psi = \psi(\cdot+a)$) satisfy 
\eq{
    \big\| \1_{\Bb_R(0)} \big(H_1^{(a)}-H_2^{(a)}\big) \1_{\Bb_R(0)} \big\| \leq \epsi.
}
Moreover, we note importantly that the definition of $\epsilon$ as $\epsilon = 2^{-10} (1+\| B_1 \|_1)^{-2}$ is translation-invariant: 
\eq{
    2^{-8} (1+\| T_a B_1 T_a^* \|_1)^{-2} = 2^{-8} (1+\| B_1 \|_1)^{-2} = \epsilon.
}
Because $T_a B_1 T_a^*$ corresponds to $B_1$ with $H_1$ replaced by $H_1^{(a)}$ and $L$ replaced by $L^{(a)}$, repeating the argument of Steps 1-4 yields $\II(H_1^{(a)}) = \II(H_2^{(a)})$. But by \Cref{lem:6}, $\II(H_j) = \II(H_j^{(a)})$. This completes the proof. 
\end{proof}

\begin{remark} The quantity $R$ produced by \Cref{prop:5} depends on $E_F$. In particular, if $E_F$ gets close to $\Spec(H_1)\cup \Spec(H_2)$, then $R$ has to be chosen large. This dependence is hidden in \textit{Step 4} above: the ``locality'' of $P_1$, $P_2$ breaks down as $E_F$ gets close to $\Spec(H_1)\cup \Spec(H_2)$. See \cite[Lemma 3.2]{DZ23} for more details. 
\end{remark}

\section{Proof of \Cref{thm:4}}\label{sec:4}

\begin{proof}[Proof of \Cref{thm:4}] Assume by contradiction that $E_F \notin \Spec(H)$. Then $H$ is a $\Theta$-invariant Hamiltonian, insulating at energy $E_F$. So is $H_+$; in particular we can apply \Cref{prop:5} to the pair $(H, H_+)$. Let $\epsi, R$ be the quantities produced by that proposition. 

Let $\rho \geq R$ such that $\nu^{-1} e^{-\nu \rho} \leq \epsi$. Let  $a \in \Z^2$ such that $\Bb_{2\rho}(a) \subset \Omega$. Then, using that $H$ and $H_+$ nearly coincide away from $\p \Omega$, in particular in the ball $\Bb_\rho(a)$: 
\eq{\label{eq:4o}
    \| \1_{\Bb_\rho(a)}(H - H_+)\1_{\Bb_\rho(a)} \| \leq \nu^{-1} e^{-\nu \rho} \leq \epsi.
} Indeed, we have \eq{
\1_{\Bb_\rho(a)}(H - H_+)\1_{\Bb_\rho(a)} = \1_{\Bb_\rho(a)}E\1_{\Bb_\rho(a)}
} where $E$ is the edge operator in \cref{eq:edge operator}. We note that $d(\Bb_\rho(a),\p\Omega) \geq d(\Bb_\rho(a),\Bb_{2\rho}(a)^c) \geq \rho$. By assumption, the kernel of $E$ decays away from $\p\Omega$, and hence for $\rho$ sufficiently large, 
\begin{equation}
    \big\| \1_{\Bb_\rho(a)}E\1_{\Bb_\rho(a)} \big\| \leq \nu^{-1} e^{-\nu \rho}.
\end{equation}
See \cite[pp. 7]{DZ23} for details. 

From \Cref{prop:5} and \eqref{eq:4o}, we deduce $\II(H) = \II(H_+)$. Likewise, $\II(H) = \II(H_-)$. This contradicts $\II(H_-) \neq \II(H_+)$, and hence completes the proof. \end{proof}

    \section{About the geometric condition on $\Omega$}\label{sec:5}
    
In this section, we prove that interface Hamiltonians between insulators that do not occupy regions containing arbitrarily large balls do not have to be conductors. Specifically, for any $L$, we construct two time-reversal invariant operators $H_\pm$ on $\ell^2(\Z^d,\C^4)$, insulating at energy $0$, such that if $\Omega \subset \Z \times [-L,L]$ then 
\eq{
    H \de H_+ + \1_{\Omega^c}(H_- - H_+) \1_{\Omega^c}
}
is still insulating at energy $0$: $0 \notin \Spec(H)$. In other words, \Cref{thm:4} fails if one relaxes the condition ``$\Omega$ and $\Omega^c$ contain arbitrarily large balls" to ``$\Omega$ and $\Omega^c$ contain a ball of radius $L$".

\subsection{Wallace and Haldane models}\label{subsec:4.1}
Our construction relies on several graphene tight-binding models.  

The Wallace model of graphene \cite{Wallace}, denoted $H_0$ below and acting on wavefunctions $\psi = (\psi^A,\psi^B) \in \ell^2(\Z^2,\C^2)$, models tunneling from each vertex to the three nearest vertices: 
 \begin{align}
 &(H_0 \psi)_n = \begin{bmatrix}
     \psi_n^B + \psi_{n - e_1}^B + \psi_{n - e_2}^B\\
     \psi_n^A + \psi_{n + e_1}^A + \psi_{n + e_2}^A
 \end{bmatrix}.
 \end{align}
It is well-known that $H_0$ has a Dirac cone at energy $0$.

We can open a gap at $0$ by perturbing $H_0$ in two different ways:
\begin{enumerate}
    \item One can add a complex second-nearest hopping $S$ below to $H_0$ to get $H_s= H_0 + s S$, $s \in R^*$ -- this is known as the Haldane model \cite{Haldane88}:
    \[
    (S\psi)_n = i \begin{bmatrix}
     \psi_{n + e_1}^A - \psi_{n + e_1 - e_2}^A + \psi_{n - e_2}^A - \psi_{n - e_1}^A + \psi_{n + e_2 - e_1}^A - \psi_{n + e_2}^A\\
     -\psi_{n + e_1}^B + \psi_{n + e_1 - e_2}^B - \psi_{n - e_2}^B + \psi_{n - e_1}^B - \psi_{n + e_2 - e_1}^B + \psi_{n + e_2}^B
 \end{bmatrix}.
    \]
    \item One can add break parity between red and blue sites by adding a directed hopping $P$ between each pair of $A$ and $B$ in a unit cell to get $W_s= H_0 + s I$, $s \in \R^*$, where
    \eq{
        (I\psi)_n = \begin{bmatrix}
     \psi_n^A\\
     -\psi_n^B
 \end{bmatrix}.
    }
\end{enumerate}
Using a Fourier transform, one can check that $0 \notin \Sigma(H_s)$ and $0\notin \Sigma(W_s)$ when $s \neq 0$; we delay the proof to \Cref{lem: spectrum at 0} in \S\ref{subsec:4.4}. Their Hall conductances are given by \cite[\S 8]{BH13}: 
\[
    c(H_s) = \operatorname{Sign}(s), \qquad c(W_s) = 0.
\]

\subsection{Construction of $H_\pm$}\label{subsec:4.2} We construct the operators $H_\pm$ by layering the Hamiltonians $H_s, W_s$ of \S\ref{subsec:4.1}. First, introduce our time-reversal operator $\Theta$ on $\ell^2(\Z^2,\C^4)$, defined by:
\eq{
   \Theta  = \matrice{0 & \Gamma \\ \Gamma & 0}, \quad \text{~where~}\quad \Gamma \psi = \begin{bmatrix}
       0 & 1\\
       -1 & 0
   \end{bmatrix}\overline{\psi}.
}
This is clearly an anti-unitary map that commutes with position, and that satisfies $\Theta^2 = -\Id$.

We now define $H_\pm$ a operator on $\ell^2(\Z^2,\C^4)$ by 
\begin{align}
   &H_+ = \begin{bmatrix}
    H_s & 0 \\ 0 & \Gamma^* H_s \Gamma
\end{bmatrix}, \qquad H_- = \begin{bmatrix}
    W_s & 0 \\ 0 & \Gamma^* W_s \Gamma
\end{bmatrix}.
\end{align}
The Hamiltonian $H_+$ is the Kane--Mele model \cite{KM05a}. Since $0\notin \Spec(H_s)$ and $0\notin \Spec(W_s)$, we have $0\notin \Spec(H_\pm)$, i.e. $H_\pm$ are insulating at energy $0$. We check that $H_\pm$ are $\Theta$-invariant:
\eq{
    H_+ \Theta = \begin{bmatrix}
    H_s & 0\\
    0 & \Gamma^* H_s \Gamma
\end{bmatrix} \begin{bmatrix}
    0 & \Gamma\\
    \Gamma & 0
\end{bmatrix} = \begin{bmatrix}
    0 & H_s \Gamma\\
    \Gamma H_s & 0
\end{bmatrix} = \begin{bmatrix}
    0 & \Gamma\\
    \Gamma & 0
\end{bmatrix}\begin{bmatrix}
    H_s & 0\\
    0 & \Gamma^* H_s \Gamma
\end{bmatrix}  = \Theta H_+,
}
with a similar calculation for $H_-$. As a result, by \cite[Prop. 4.11]{FSSWY20}, we have 
\[
\II(H_+) = c(H_s) = 1 \mod 2, \quad \II(H_-) = c(W_s) = 0 \mod 2. 
\]
Hence $H_\pm$ are two $\Theta$-invariant Hamiltonians with $0\notin \Spec(H_\pm)$ and $\II(H_+) \neq \II(H_-)$. 

\subsection{Construction of $H_e$}\label{subsec:4.3}
We define $H_e$ on $\ell^2(\Z^2,\C^4)$ by: 
\[
H_e = H_+ + \1_{\Omega^c}(H_- - H_+) \1_{\Omega^c}.
\]
Note $E= H_e - \1_{\Omega}H_+\1_{\Omega} - \1_{\Omega^c} H_-\1_{\Omega^c} = \1_{\Omega} H_+\1_{\Omega^c} + \1_{\Omega^c} H_+ \1_{\Omega}$, by \cite[Lemma 2.2, (2.3)]{DZ24}, we have 
\[
|E(x,y)|\leq Ce^{-d(x, \partial \Omega) }.
\]
Hence $H_e$ is an interface system in the sense of \Cref{def:edge} -- implicitly depending on the parameter $s$. However:

\begin{proposition}
    Assume $\Omega\subset \Z\times [-L, L]$ for some $L > 0$. Then there exists $s>0$ such that $H_e$ is invertible.
\end{proposition}

\begin{proof}
    Note that $H_- - H_+ = s \cdot \begin{bmatrix}
        I - S & 0 \\
        0 & \Gamma^* (I - S) \Gamma
    \end{bmatrix} =: s \Delta$ and hence 
    \[
    H_e = H_+ \cdot \big( \Id -  s \cdot H_+^{-1}\1_{\Omega^c}\Delta\1_{\Omega^c}\big)
    \]
    To prove that $H_e$ is invertible, it suffices now to show that $\Vert mH_+^{-1}\1_{\Omega^c}\Delta\1_{\Omega^c}\Vert <1$ for some $s > 0$, and invoke a Neuman series argument. In fact, by \cite[Lemma 4.2]{DZ23}, there is $C_0>0$, such that for all $m\in (0, 1]$, $L>0$, and $u\in \ell^2(\Z^2,\C^4)$, we have 
    \[
    \supp(u)\subset \Z^2\times [-L, L] \qquad \Rightarrow \qquad \Vert H_+^{-1}u\Vert \leq C_0 L^{1/3} s^{-2/3} \Vert u \Vert.
    \]
    As a result, since $\Vert \Delta \Vert \leq 8$, we have 
    \[
    \Vert s \cdot H_+^{-1}\1_{\Omega^c}\Delta\1_{\Omega^c}\Vert\leq C_0 L^{1/3} s^{1/3} \Vert \1_{\Omega^c} \Delta\1_{\Omega^c} \Vert \leq 8C_0 L^{1/3} s^{1/3}  <1 
    \]
    when $s$ is small enough. This completes the proof.  
\end{proof}

\subsection{Spectrum at $0$}\label{subsec:4.4}
\begin{lemma}\label{lem: spectrum at 0}
    If $m \neq 0$, we have $0\notin \Spec(H_s)$, $0\notin \Spec(W_s)$, and $0\notin \Spec(H_\pm)$.
\end{lemma}

\begin{proof}
    Indeed, let $U$ denote the Fourier transform $U:\ell^2(\Z^2, \C^2)\to L^2([-\pi, \pi]^2, \C^2)$ by 
\[
\hat\psi(\xi)\de (U\psi)(\xi)\de \sum\limits_{n\in \Z^2} e^{-i\langle n, \xi\rangle} \psi_n.
\]
Then $\hat{H}_0 = UH_0U^*$ is a matrix-valued multiplication operator from $L^2([-\pi, \pi]^2, \C^2)$ to itself:
\[
\hat{H}_0(\xi)= (UH_0U^*)(\xi) = \begin{bmatrix}
    0 & \overline{\omega(\xi)}\\
    \omega(\xi), & 0
\end{bmatrix}, \quad \text{where} \quad \omega(\xi) = 1 + e^{i\xi_1} + e^{i\xi_2}. 
\]
Since eigenvalues of $\hat{H}_0(\xi)$ are $\lambda_0(\xi) = \pm |\omega(\xi)|$, we have 
\[
\Spec(H_0) = \bigcup_{\xi\in [-\pi, \pi]^2} \Spec(\hat{H}_0(\xi)) = \bigcup_{\xi\in [-\pi, \pi]^2} \{\pm|\omega(\xi)|\}.
\]
Because $\omega(\xi) = 0$ when $\xi = \xi_\pm = \pm \frac{2\pi }{3}(1, -1)$, we have $0\in \Spec(H_0)$. 

Note that $S$ models the directed second-nearest hopping while $B$ breaks the parity between sites $A$ and $B$. A simple calculation gives
\[
(\hat{H}_s)(\xi) = \begin{bmatrix}
    \pm2m\eta(\xi) & \overline{\omega(\xi)}\\
    \omega(\xi), & \mp 2m \eta(\xi)
\end{bmatrix}, \qquad (\hat{W}_s)(\xi) = \begin{bmatrix}
    \pm1 & \overline{\omega(\xi)}\\
    \omega(\xi), & \mp1
\end{bmatrix}
\]
where $\eta(\xi) = \sin(\xi_1) - \sin(\xi_2) + \sin(\xi_2 - \xi_1)$. Hence eigenvalues of $(\hat{H}_s)(\xi)$ and $(\hat{W}_s)(\xi)$ are $\lambda_1(\xi) = \pm \sqrt{(2s\eta(\xi))^2 + |\omega(\xi)|^2}\neq 0$ for all $\xi$ and $\lambda_2(\xi) = \pm \sqrt{1 + |\omega(\xi)|^2}\neq 0$ for all $\xi$. As a result, $0\notin \Spec(H_s) = \Spec(H_+)$ and $0\notin \Spec(W_s) = \Spec(H_-)$.
\end{proof}

\bibliographystyle{amsxport}
\bibliography{ref.bib}

\end{document}